\documentclass[10pt]{iopart}

\usepackage{epsf}
\newcommand \be {\begin{equation}}
\newcommand \bea {\begin{eqnarray}}
\newcommand \ee {\end{equation}}
\newcommand \eea {\end{eqnarray}}
\newcommand \s {\sigma}

\begin{document}

\title{Dynamical phase transitions in glasses induced by the
ruggedness of the free energy landscape}

\author{F. Ritort}

\address{\dag\ Physics Department, Faculty of Physics \\
         University of Barcelona, Diagonal 647, 08028 Barcelona,  Spain
        }

\begin{abstract}
We propose damage spreading (DS) as a tool to investigate the
topological features related to the ruggedness of the free energy
landscape. We argue that DS measures the positiveness of the largest
Lyapunov exponent associated to the basins of attraction visited by
the system during its dynamical evolution. We discuss recent results
obtained in the framework of mode-coupling theory and comment how to
extend them to the study of realistic glasses. Preliminary results are
presented for purely repulsive soft-sphere glasses.
\end{abstract}


\maketitle

\section*{Introduction}
The theoretical understanding of a first principles theory for the
glass transition is still missing. Despite of great advances in the
understanding of some generic features associated to the glass
transition (such as those predicted by the mode-coupling
theory) still some questions remain largely unknown. Going beyond the
schematic mode-coupling theory seems to be an enormous task so an
alternative way of looking at the glass transition may be useful.  In
this direction, the study of the topological properties of both the
potential or free energy landscape may yield further information on
the mechanisms responsible for the anomalous viscosity of the glassy
phase.

The idea that topological aspects of the potential or free energy
landscape are the ultimate reason for the glass transition goes back to
Goldstein \cite{GO} and more recently Stillinger and Weber
\cite{SW,SW2}. This approach has been recently applied to the study of
hard-spheres \cite{DV}, monoatomic as well as binary Lennard-Jones glasses
\cite{SH} or mean-field models of glasses \cite{CR1}.

Here we propose an alternative dynamical approach to study the
topological properties of the potential energy landscape. We will
concentrate on the study of the stability local properties of the
configurations visited by the system during its dynamical
evolution. This is directly achieved through the study of how dynamical
trajectories, which evolve following the same stochastic noise, depart
from each other in the presence of a potential energy saddle point or a
maximum which may induce a negative Lyapunov exponent.  The simplest way
to study this problem is through damage spreading (DS) techniques to be
describe later on in some detail.  Although DS was introduced almost two
decades ago as an alternative way to consider thermodynamic phase
transitions, the initial enthusiasm on this problem strongly decayed
when it was realized that DS transitions are not universal and not
necessarily related to thermodynamic singularities.

Despite of this result here we will show that these transitions have
an added interest in that they may be used as a direct way to
investigate the local free energy landscape properties by measuring
the largest Lyapunov exponent associated with the Hamming distance (to
be defined later). In what follows I will explain in more detail why
DS is a good way of looking at the rugged properties of the potential
energy landscape. Later on I will discuss the analytical results
obtained for the schematic mode-coupling theory and finally discuss
how to extend these ideas to the study of real glasses. Some
preliminary results are shown for the case of binary soft-sphere
purely repulsive glasses.

\section{Why damage spreading?}

Consider two systems evolving under a Langevin dynamics each one
described by a set of $N$ variables $x_i,y_i;1\le i\le N$ evolving in
a potential energy landscape ${\cal V}$ under the same stochastic noise
$\eta_t$ with $\langle
\eta_i(t)\eta_j(s)\rangle=2T\delta_{ij}\delta(t-s)$. Although the
present discussion can be generalized for different stochastic noises
here we will concentrate on the simplest case (for a more detailed
discussion see \cite{HR}). The equations of motion read,

\bea
\dot{x}_i(t)=F_i(\lbrace x\rbrace)+\eta_i(t)\label{1}\\
\dot{y}_i(t)=F_i(\lbrace y\rbrace)+\eta_i(t)\label{2}
\eea

\noindent
where $F_i(\lbrace x\rbrace)=-\frac{\partial {\cal V}}{\partial x_i}$.
Note that both trajectories described by the systems $x$ and $y$ never
cross in phase space so two identical configurations such that
$x_i(t)=y_i(t)$ remain identical forever (and where identical in the
past). The equation for the difference variables $z_i=x_i-y_i$ reads,

\be
\dot{z}_i(t)=F_i(\lbrace x_i\rbrace)-F_i(\lbrace y_i\rbrace)~~~.
\label{3a}
\ee

\noindent
If the $z_i$ are small we can expand (\ref{3a}) around $z_i=0$
obtaining,

\be
\dot{z}_i(t)=\sum_j \frac{\partial F_i(\lbrace y\rbrace)}{\partial
y_j}z_j=-\sum_j \frac{\partial^2 V(\lbrace y\rbrace)}{\partial
y_i\partial y_j}z_j \label{3b}
\ee

\noindent
which may be written in a simplified form,

\be
\dot{z}_i=H_{ij}(\lbrace y\rbrace)z_j
\label{4}
\ee

\noindent
where $H_{ij}$ is the Hessian matrix evaluated at the configuration
$y$. Always within the linear approximation the dynamical evolution of
the distance between configurations $z_i$ will increase or decrease
according whether the spectrum of eigenvalues of the Hessian matrix
contains positive eigenvalues. In this sense, DS probes the spectrum of
eigenvalues of the matrix and shows instabilities whenever the matrix
develops positive eigenvalues. A more precise condition is given by the
maximum Lyapunov exponent defined through,

\be
\lambda_{max}=\lim_{t\to\infty}\frac{\log(D(t))}{t}
\label{5}
\ee

where $D(t)=\frac{1}{N}\sum_i z_i^2$ which should be positive whenever
$z_i=0$ is dynamically unstable. Note that the Hessian depends on time
through the time evolution of the generic configuration $y$. This may be
an equilibrium or an off-equilibrium configuration. So in principle the
maximum Lyapunov exponent depends on time through the time evolution of
the systems $x$ and $y$. We will see later that, in general, the type of initial
condition (as well as the initial distance) are not relevant parameters
for the DS transition. In this sense DS probes the temperature at which
the lowest accessible configurations in the potential energy landscape
develop unstable modes being a direct check of the corrugated properties
of the free energy landscape. Again, we must insist on the
non-universal properties of the DS dynamics. The present discussion
on the stability properties of the Hessian matrix and its connection
with the DS transition is valid in the framework of Langevin
dynamics. For other type of dynamics (such as Monte Carlo
dynamics or Glauber) the situation may be different and the physical
meaning of DS phenomena more difficult. In some sense, Langevin dynamics
is an appropriate tool to explore the topological properties of the
potential energy landscape.

\section{DS in mode-coupling theory}

Insight on the previous problem can be obtained through a careful study
of the DS equations in the case of ideal mode-coupling theory. It is
known since the seminal work by Kirkpatrick, Thirumalai and Wolyness
\cite{KTW} that mode coupling equations can be obtained in the framework
of exactly solvable $p$-spin glass models. Due to their mean-field
character, in this class of models it is possible to unambiguously define
concepts such as the configurational entropy or complexity and the
mode-coupling transition temperature $T_c$. The description of this type
of models is possible in the framework of the TAP analysis \cite{CS}
where it is possible to show that they contain a large number of
metastable states (exponentially large with $N$) as
well as a threshold energy where the system gets trapped in an
asymptotic aging state and the fluctuation-dissipation theorem is
violated in a peculiar way \cite{BCKM}.

Spherical p-spin models (contrarily to Ising spins) have the clear
advantage of being exactly solvable so it is convenient to do analytical
computations in that case. The potential energy in this model is defined
by, 

\be
{\cal V}=-\sum_{(i_1<i_2<...<i_p)}\,J_{i_1,i_2,i_3,..,i_p}\s_{i_1}\s_{i_2}\s_{i_3}..\s_{i_p}~~~.
\label{6}
\ee

where the spins $\s_i$ are real valued spins which satisfy the spherical
constraint $\sum_{i=1}^N\s_i^2=N$. The
$J_{i_1,i_2,i_3,..,i_p}$ are quenched random variables with zero mean and
variance $p!/(2N^{p-1})$. The Langevin dynamics of the model is given by,

\be
\frac{\partial \s_i}{\partial t}=F_i(\lbrace\s\rbrace)-\mu\s_i+\eta_i
\label{7}
\ee

\noindent
where $\mu$ is a Lagrange multiplier which ensures that the spherical
constraint is satisfied at all times and the noise $\eta$ satisfies the
fluctuation-dissipation relation $\langle\eta_i(t)\eta_j(s)\rangle=2T
\delta(t-s)\delta_{ij}$ where $\langle...\rangle$ denotes the noise
average.  $F_i$ is the force acting on the spin $\s_i$ due to the
interaction with the rest of the spins,

\be
F_i=-\frac{\partial V}{\partial s_i}=\frac{1}{(p-1)!}
\sum_{(i_2,i_3,...,i_p)}\,J_{i_1,i_2,..,i_p}\s_{i_2}\s_{i_3}..\s_{i_p}~~~.
\label{8}
\ee

We define the overlap between two configurations of the spins
$\s,\tau$ by the relation $Q=\frac{1}{N}\sum_{i=1}^N\s_i\tau_i$ so
a distance between these two configurations is,

\be
D=\frac{1-Q}{2}
\label{9}
\ee

\noindent
in such a way that identical configurations have zero distance and
opposite configurations have maximal distance $D=1$. Then we consider
two copies of the system $\lbrace\s_i,\tau_i\rbrace$ which evolve under
equation (\ref{7}) with the same statistical noise and start from random
initial configurations.

The final equations are \cite{HR},

\bea
\hspace{-2truecm}
\frac{\partial C(t,s)}{\partial t}+\mu(t)C(t,s)=
\frac{p}{2}\int_0^s du R(s,u)C^{p-1}(t,u)+\nonumber\\
\frac{p(p-1)}{2}
\int_0^t du R(t,u)C(s,u)C^{p-2}(t,u)\label{10}\\
\hspace{-2truecm}
\frac{\partial R(t,s)}{\partial t}+\mu(t)R(t,s)=\delta(t-s)
+\frac{p(p-1)}{2}
\int_s^t du R(t,u)R(u,s)C^{p-2}(t,u)\label{11}\\
\hspace{-2truecm}
\frac{\partial Q(t,s)}{\partial t}+\mu(t)Q(t,s)=
\frac{p}{2}\int_0^s du R(s,u)Q^{p-1}(t,u)
\nonumber\\+\frac{p(p-1)}{2}
\int_0^t du R(t,u)Q(u,s)C^{p-2}(t,u)\label{12}
\eea

\noindent
The dynamical equations involve the two times correlation, response
and overlap function $C(t,s), R(t,s), Q(t,s)$ defined by (in what
follows we take $t>s$),

\bea
C(t,s)=(1/N)\sum_{i=1}^N\langle\s_i(t)\s_i(s)\rangle=(1/N)\sum_{i=1}^N\langle\tau_i(t)\tau_i(s)\rangle
\label{13}\\
R(t,s)=(1/N)\sum_{i=1}^N\frac{\partial\langle\s_i\rangle}{\partial h^{\s}_i}=
(1/N)\sum_{i=1}^N\frac{\partial\langle\tau_i\rangle}{\partial h^{\tau}_i}\label{14}\\
Q(t,s)=(1/N)\sum_{i=1}^N\langle\s_i(t)\tau_i(s)\rangle\label{15}
\eea

\noindent
where $<..>$ denotes the average over dynamical histories and
$h^{\s}_i,h^{\tau}_i$ are fields coupled to the spins $\s_i,\tau_i$
respectively. These equations are complemented with the appropriate
boundary conditions $C(t,t)=1, Q_d(t)=Q(t,t), R(s,t)=0, \lim_{t\to
(s)^+} R(t,s)=1$ and the relations,

\bea
\mu(t)=T+\frac{p^2}{2}\int_0^tdu R(t,u)C^{p-1}(t,u)\label{16}\\
\frac{1}{2}\frac{\partial Q_d(t)}{\partial t}+\mu(t)Q_d(t)=T+
\frac{p}{2}\int_0^t du R(t,u)Q^{p-1}(t,u)
\nonumber\\
+\frac{p(p-1)}{2}\int_0^t du R(t,u)Q(t,u)C^{p-2}(t,u)\label{17}~~~~~~.
\eea

These equations can be analyzed in detail 
using different methods. Here we summarize the main results
obtained \cite{HR},

\begin{itemize} 

\item{Existence of a dynamical transition $T_0$}

There is a temperature $T_0$ such that $D(t)=0$ (or $Q_d(t)=1$, see eq.(\ref{9}))
is a stable fixed point for $T>T_0$ becoming unstable below $T_0$.
Because of the non-monotonic character of $D(t)$ it is very difficult to
derive analytically $T_0$. Nevertheless, it is possible to obtain an upper
and a lower bound. One gets,

\be
\sqrt{\frac{p-2}{2}} \le T_0\le \sqrt{\frac{p}{2}}
\label{18}
\ee

Direct numerical integration of the equations of motion yields
$T_0(p=3)=1.04 \pm 0.02$ with and $T_0(p=4)=1.13 \pm 0.02$. The value of
$T_0$ is well above the mode-coupling temperature $T_c$ and the TAP
temperature $T_{TAP}$ below which there is an exponentially large (with
the system size) number of metastable states.

\begin{figure}[hbt]
\epsfxsize=7cm\epsfysize=7cm \epsfbox{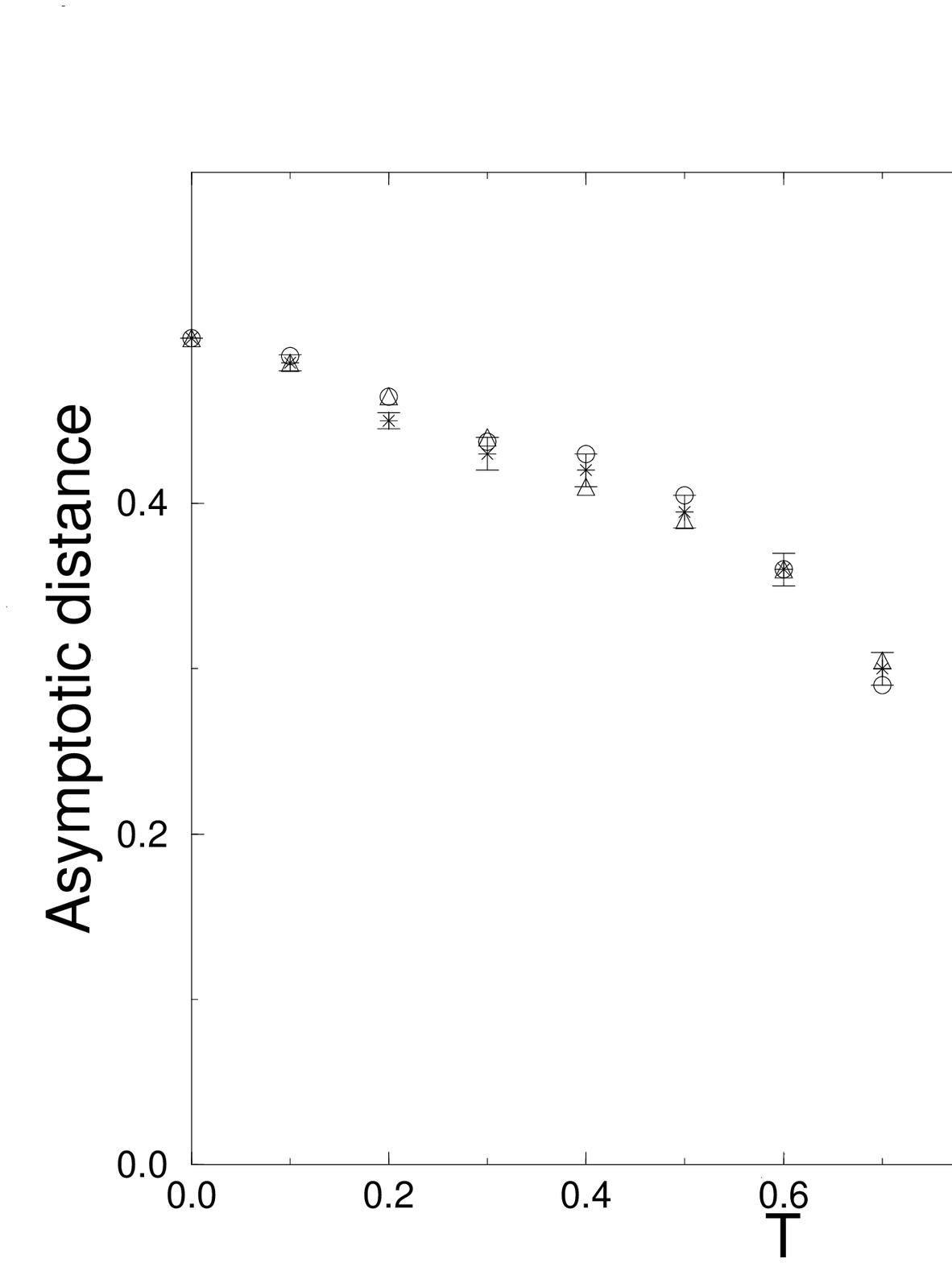}
\caption{Asymptotic distance $D_{\infty}$ for $p=3$
($\alpha=1,{\cal K}=1$) obtained from the Pade analysis of the series
expansions for different initial conditions $D_0=1$ (circles), $D_0=0.5$
(triangles), $D_0=0.25$ (stars). Typical error bars are shown for the
last case.
}
\label{fig1}
\end{figure}

\item{Independence of initial conditions}

The asymptotic damage $D(\infty)=\lim_{t\to\infty}\lim_{N\to\infty}
D(t)$ is independent on the value of the initial damage $D(0)$ or the
class of initial conditions (for instance, random or
thermalized). This independence stresses the fact that DS is a true
dynamical transition and the asymptotic damage $D(\infty)$ is a
dynamical order parameter.

\begin{figure}[hbt]
\epsfxsize=7cm\epsfysize=7cm \epsfbox{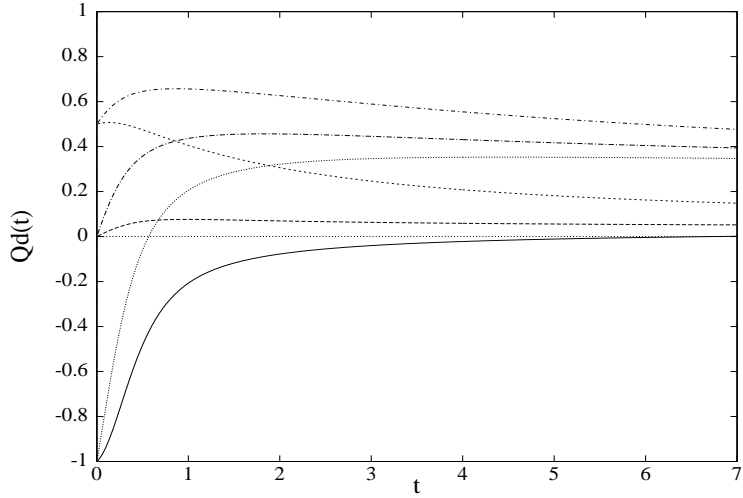}
\caption{$Q_d(t)$ for $p=3$ ($\alpha=1,{\cal K}=1$) at
  temperatures $T=0.1,0.5$ (from bottom to top at large times) for three
  different values of the initial overlap $Q_d(0)=-1,0,0.5$ as a
  function of time. The continuous lines are the numerical integrations
  with time step $\Delta t=0.01$.
}
\label{fig2}
\end{figure}

\item{$T_0$ is the lowest DS temperature} 

The DS problem can be suitably generalized for the case of correlated
noises such that $\langle\eta_i(t)\xi_j(s)\rangle=2T {\cal
K}(Q(t,s))\delta(t-s)\delta_{ij}$ where $\eta$ and $\xi$ are the noises
acting on the systems $\s$ and $\tau$ respectively. The function ${\cal
K}$ satisfies ${\cal K}(1)=1$ so both noises are identical if the two
configurations coincide. This implies that $Q_d(t)=1$ is a fixed point
of the dynamics. It can be shown that for any possible function ${\cal
K}\le 1$ (with ${\cal K}(1)=1$) there is a finite temperature damage
spreading transition $T_0$ only if ${\cal K'}(1)\le 1$. The case
discussed previously ${\cal K}=1$ (identical noises at all times) yields
the lowest damage spreading transition temperature. 

\item{$T_0$ is the endpoint of a dynamical critical line}

The DS problem can be also generalized to the case ${\cal
K}(Q)\le\lambda$ with $\lambda\le 1$ and ${\cal
K}(1)=\lambda$. Obviously for $\lambda=1$ dynamical trajectories of
both systems may cross. In this case it is possible to show that the
function ${\cal K}(Q)=\lambda$ yields the lowest DS transition
temperature $T_0(\lambda)$ among the set of possible functions ${\cal
K}$ (${\cal K}(Q)\le\lambda, {\cal K}(1)=\lambda$). $T_0(\lambda)$ is
monotonic increasing function function of $\lambda$ which for
$\lambda=0$ coincides with the mode-coupling transition temperature
$T_c$ and finishes in a critical endpoint $T_0(\lambda=1)=T_0$. So
there exists a line of dynamic critical points which connect the
mode-coupling temperature $T_c$ with the DS temperature $T_0$.

\item{$T_0$ is not universal.} 

The temperature $T_0$ is not universal. As it depends on the set of
correlations of the noises it also depends on the type of dynamics
(molecular dynamics, Monte Carlo with Metropolis, heat-bath or Glauber).
This is a well known result which finds its natural explanation on the
physical origin of the DS transition. For a general dynamics it is not
possible to map the DS transition with the local properties of the
potential energy landscape. Only for the case of Langevin dynamics or
molecular dynamics this is possible. Other dynamics (such as Monte Carlo
with heat-bath dynamics) use random numbers in the dynamics which
introduce complex correlations between the noises. This yields a DS
transition (related to the $T_0(\lambda)$ discussed in the previous
paragraph for the Langevin case) which is probably related with the
mode-coupling transition temperature but this issue still needs to be
further investigated.

\end{itemize}

\section{Application to binary soft-sphere glasses}

In this section we apply the previous ideas derived in the framework of
mode-coupling theory to the case of structural glasses. We consider the
binary soft-spheres model introduced in \cite{SP} and recently studied
in \cite{CMPV}. For sake of simplicity we consider a gas of $N$
particles such that half of them have diameter $\sigma_1$ and the other
half $\sigma_2$. The particles interact through a two particle purely
repulsive potential, the energy of the system being defined by

\be
{\cal V}=\sum_{i<j}\bigl ( \frac{\sigma_{ij}}{r_{ij}}\bigr )^{12}
\label{19}
\ee

The choice $\sigma_{ij}=\frac{\sigma_{i}+\s_{j}}{2}$ supposes that
diameters are additive during the collision process. The advantage of
this potential is that the thermodynamic properties depend on the
density $\rho=N/V$ and the temperature $T$ only through the constant
$\Gamma=\rho/T^{\frac{1}{4}}$. For the particular case
$\frac{\s_1}{\s_2}=1.2$ crystallization is strongly inhibited and the
glass transition (where dynamics is strongly slowed down) appears in
the vicinity of $\Gamma=1.45$. Larger values of $\Gamma$ correspond to
the glass phase while lower values correspond to the liquid phase.
The Langevin dynamics for the soft-sphere model is defined by,

\be
\dot{\vec{r}_i}=-\sum_{j\ne i}^N \nabla_iV_{ij}(r_{ij})+\vec{\eta_i}
\label{20}
\ee

\noindent
with
$\langle\eta_i^k(t)\eta_j^l(t')\rangle=2T\delta_{ij}\delta_{kl}\delta(t-t')$
where the superindex in the noise indicate the different Cartesian
components of the vector noise $\vec{\eta}(t)$. The pairwise potential
is given by $V_{ij}(r)=(\frac{\s_{ij}}{r})^{12}$. 

We now consider two systems described by the variables
$\vec{r_i},\vec{s_i}$ governed by (\ref{20}) and evolving under the same
realization of the noise. We define the Euclidean distance 

\be
D(t)=\frac{1}{N}\sum_{i=1}^N(\vec{r_i}-\vec{s_i})^2
\label{21}
\ee

which vanishes if the two configurations coincide. If we want to extend
the previous ideas for the spherical $p$-spin model to this system now we
must take into account the fact that at very high temperatures a gas
diffuses so $D=0$ may not be a fixed point of the dynamics. 
There are two strategies to deal with this problem which are discussed below.

\begin{itemize}

\item{Particles contained in a box}

This is the most natural choice.  To simulate a purely repulsive system
one must confine the particles in a cubic box of side L such that
$\rho=\frac{N}{L^3}$. In this case one may numerically solve (\ref{19})
with two different class of boundary conditions. With periodic boundary
conditions particles leave one side of the box and enter the opposite
side. This resets completely the coordinates of the particle so the
distance (\ref{21}) is discontinuous if particles cross the
boundaries. Concerning one system quantities (such as the energy or the
pair correlation function) this is not a problem because the relevant
quantity is the distance between the particles which may be taken as the
minimum value between $r_{ij}$ and $L-r_{ij}$. A similar procedure can
be used to define the distance between the two copies. Everything can be
easily solved considering free boundary conditions so particles are not
allowed to cross the boundaries. In this case, it is possible to show
that $D=0$ is asymptotically stable for the purely diffusive case
($\Gamma=0$).

Preliminary results show that the DS transition temperature $T_0=\infty$
so two configurations never coincide at finite temperature. Still both
configurations retain some correlation (so $\langle
\vec{r_i}(t)\vec{s_i}(t)\rangle > 0$) and the asymptotic damage is a
non trivial function of the temperature.

\begin{figure}
\epsfxsize=7cm\epsfysize=7cm \epsfbox{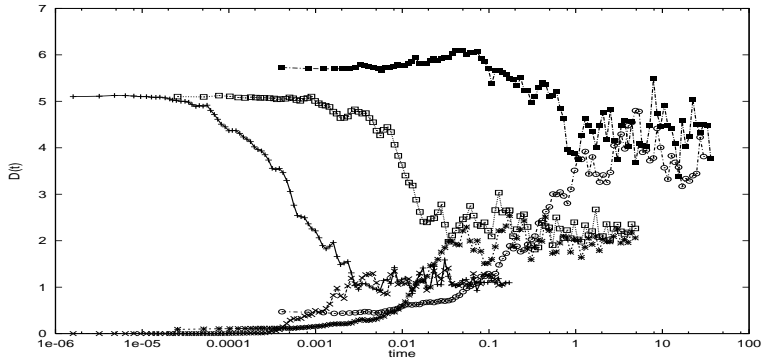}
\caption{Damage $D(t)$ as a function of time for $N=32$ starting from
two different initial conditions and three different temperatures
(from top to bottom)  $\Gamma=0.8,0.4,0.2$}
\label{fig3}
\end{figure}

\item{Introducing an spherical constraint}

For the purpose of studying the local properties of the potential energy
landscape we may impose the global constraint $\sum_i
\vec{r_i}^2=N(\frac{N}{\rho})^{\frac{2}{3}}$ on the particles in such a
way that the average distance between the particles is finite when $N$
goes to infinity.  Because the
spherical constraint shifts the Hessian matrix (\ref{4}) by a constant
(a Lagrange multiplier) the transition with the spherical constraint
may give information on the transition for the unconstrained
case. That Lagrange multiplier can be simply obtained from the potential
energy $<{\cal V}>$ and the temperature.
The advantage of such a constraint is that now there
is no box and $D=0$ is a fixed point of the dynamics for $\Gamma=0$. The
inconvenience is that the simplicity of the original model is lost and
the thermodynamics of the new model depends on both density and
temperature instead of a unique parameter $\Gamma$.

Again, preliminary results show that $T_0=\infty$ in this case so
$D=0$ is asymptotically stable strictly only for $\Gamma=0$. Although this
approach is more involved it is probably the best way to relate the DS
transition to the ruggedness of the free energy landscape.

\end{itemize}

\section{Conclusions}

The study of the free energy landscape may yield valuable information on
the glass transition phenomena. A promising description of the glass
transition is through the Stillinger and Weber projection of the
partition function in terms on inherent structures. That method directly
looks at the potential energy landscape described in terms of basins of
attraction explored by the system during its dynamical evolution
\cite{CR1}. An alternative approach studies the dynamical properties of
the free energy landscape directly looking at the largest Lyapunov
exponent of the Hessian matrix of the potential energy landscape
weighted by the size of the basins of attraction visited by the system
during its dynamical evolution. 

Exact results for the mode coupling theory reveal that there is a
transition $T_0$ which separates two well defined regime depending on
the value of the asymptotic distance. Below $T_0$ the asymptotic damage
is non zero and independent of the initial distance as well as the class
of initial conditions. Above $T_0$ the damage vanishes. We argue that
the precise value of $T_0$ is related to the vanishing of the largest
Lyapunov exponent defined in (\ref{5}). Although such an explicit
connection needs still to be done it is quite probable that DS is a
precise tool to investigate the chaotic properties of the free energy
landscape. A result in this direction has been recently obtained by
Biroli through the study of the instantaneous normal modes spectra of
the $p$-spin model \cite{BIROLI}. Whether this transition has
experimental relevance in the study of real glasses is still an open
question. Our preliminary studies of soft-sphere binary mixtures show
that $T_0$ is extremely large. Because liquids are always diffusive at
large temperatures (a feature which is directly encoded in the
wave-vector dependence of correlation functions, a general feature of
liquids) one must be careful when extending the results obtained for the
spherical $p$-spin model to real structural models of glasses.  Although
a better understanding of the extension of DS to diffusive systems is
needed we can point out other interesting open problems.  One the one
hand it could be very interesting to analyze the DS transition for
molecular dynamics. In that case, there is no stochasticity in the
dynamical equations so the {\em effective} source of noise comes out
directly from the mixing property of the dynamics. The analog of
equation (\ref{4}) should be very similar except for the presence of
oscillations. Still the general argument would be the same and $T_0$
expected to be identical. Such an analysis would be
welcome. Finally it would be very interesting to look at the other
endpoint of the dynamic critical line. Our present discussion was
centered on the case of identical noises. For completely uncorrelated
noises the dynamical transition temperature is expected to coincide with
the mode-coupling transition temperature. This is true in the framework
of the aforementioned exact calculations in the spherical model and
could be also analyzed for real glasses.

\ack I warmly thank G. Biroli, A. Crisanti, S. Franz, M. Heerema,
J. Kurchan and I. Pagonabarraga for useful discussions. This work has
been supported by the Spanish Ministery of Education (PB97-0971).


\end{document}